\documentclass[12pt]{article}
\usepackage{amssymb,amsmath,epsfig}

\begin{document}

\title{\bf Cosmological Evolution for Dark Energy Models in $f(T)$ Gravity}
\author{M. Sharif \thanks {msharif.math@pu.edu.pk} and Sehrish Azeem
\thanks{sehrishazeem4@gmail.com}\\
Department of Mathematics, University of the Punjab,\\
Quaid-e-Azam Campus, Lahore-54590, Pakistan.}

\date{}
\maketitle

\begin{abstract}
In this paper, we investigate the behavior of equation of state
parameter and energy density for dark energy in the framework of
$f(T)$ gravity. For this purpose, we use anisotropic LRS Bianchi
type I universe model. The behavior of accelerating universe is
discussed for some well-known $f(T)$ models. It is found that the
universe takes a transition between phantom and non-phantom phases
for $f(T)$ models except exponential and logarithmic models. We
conclude that our results are relativity analogous to the results of
FRW universe.
\end{abstract}
{\bf Keywords:} $f(T)$ gravity; LRS Bianchi type I universe;
Equation of state; Dark energy.\\
{\bf PACS:} 04.50.kd; 98.80.-k.

\section{Introduction}

The extensions of general relativity seem attractive to explain the
late time acceleration of the universe and dark energy (DE). High
redshift type Supernova Ia experiments show that the universe is
experiencing accelerated expansion in every direction, while the
other observations of anisotropies give indirect evidence
(Perlmutter et al. 1999; Knop et al. 2003; Riess et al. 1998). The
mysterious anti-gravity DE material is smoothly filled in the
universe. Dark energy with negative pressure and positive energy
density depends on the equation of state (EoS), $p=\omega\rho$,
where $\omega$ is the function of cosmic time called EoS parameter
(Sharif and Zubair 2010a). There are different forms of dynamically
varying DE phases related with negative behavior of EoS parameter.
The range, $-1<\omega<-1/3$, corresponds to the non-phantom phase of
the universe while the phantom phase occurs for $\omega<-1$ (Yadav
2011; Sahni and Starobinsky 2008). Quintom is such a DE model which
can cross the phantom divide line $\omega=-1$ from both sides
(Khatua et al. 2011; Feng et al. 2005).

The dynamical nature of DE can originate from a variable
cosmological constant, phantom as well as scalar field, tachyon,
Chaplygin gas and modified gravities (Martinelli and Melchiorri
2009). In spite of all observational evidences, the expanding
universe is still a challenging issue in modern physics (Yang et al.
2010). An alternative approach to accommodate the current
accelerating expansion of the universe is to modify the GR on large
scale, such as, the scalar-tensor theories, $f(R)$ theory, $f(T)$
theory and DGP braneword theory (Yang 2011a; Zheng and Huang 2011)
etc.

Among these theories, the generalized teleparallel theory of gravity
has recently gained a lot of interest due to its possible
explanation about DE. In this generalization, the torsion scalar,
$T$, is replaced by its general function $f(T)$ in the Lagrangian of
teleprallel gravity (Sharif and Rani 2011a). The $f(T)$ gravity
models use the Weitzenb$\ddot{o}$ck connection which inherits only
torsion and responsible for the accelerating expansion of the
universe (Myrzakulov 2011). This approach (theory) originally
developed by Einstein in $1928$ under the name "Fern-Parallelismus"
or "distant Parallelism" or "teleparallelism" (Linder 2010; Unzicker
and Case 2005). The $f(T)$ theory has significant advantage of its
second order field equations as compared to $f(R)$ theory with
fourth order field equations (Sharif and Shamir 2009; Sharif and
Kausar 2010, 2011a, 2011b; Jamil et al. 2012a, 2012b).

Bengochea and Ferraro (2009) described the recently detected
acceleration of the universe without DE and performed observational
viability tests by using recent SNIa data for some $f(T)$ models.
Linder (2010) explored that the cosmological constant is not only
the component for the acceleration of the universe. This was also
observed through a generalization of GR to other gravity theories.
Zhang et al. (2011) discussed the dynamical analysis for the
logarithmic and power form of $f(T)$ models. They investigated
phase-space analysis of models, their stability of the critical
points and evolution of EoS parameter. In a recent paper, Sharif and
Shamaila (2011b) constructed some $f(T)$ models by using Bianchi type
I (BI) universe. They also derived EoS parameter for two modified
teleparallel models. Yang (2011b) investigated three new $f(T)$
models and described their physical implications and cosmological
behavior.

Recently, Bamba et al. (2011) discussed different $f(T)$ models to
investigate the cosmological evolutions of EoS parameter for DE and
their observational constraints. Wu and Yu (2011) proposed two new
$f(T)$ models and analyzed that the crossing of the phantom divide
line is consistent with recent cosmological observational data. Chen
et al. (2011) investigated this problem by extending modified
gravity at the background and perturbed level. They also explored
this theory for quintessence scenario.

One of the predictions from inflation is that the observed universe
should nearly isotropic on large scales. Spatially homogenous and
isotropic universe can be well described by FRW model which can not
be explained the early universe. The existence of anisotropy at
early times is a natural phenomenon to investigate the local
anisotropies that we observe today in galaxies and cluster. Recent
measurements on the cosmic microwave background radiation (CMBR)
through WMAP suggest that the anisotropic inflationary models should
be considered (Vielva 2004; Costa 2004). In early epoch of the big
bang, the universe does not maintain its isotropic behavior at very
small scales (Kumar and Singh 2008). In order to get a realistic
model representing an expanding, homogenous and anistropic universe,
Bianchi type cosmological models are considered. The isotropic
behavior of today universe makes BI model a prime candidate for
describing the possible anisotropic effects of the early universe on
modern day data observations.

Many people worked on simplest anisotropic model called BI universe
to discuss the effects of anisotropy in several contexts. Kumar and
Singh (2007) developed mechanism for solution of the field equations
by considering BI universe model. The same authors (Kumar and Singh
2008) investigated the exact BI solutions which give the constant
value of the deceleration parameter in scalar-tensor theory.
Recently, Sharif and Waheed (2012) studied exact solutions for
anisotropic fluids by taking the locally rotationally symmetric
Bianchi Type I (LRS BI) universe model which generalizes the flat
FRW universe in the modified theory (BD). There are many papers
available in literature where this model has been used widely (Bali
and Kumawat 2008; Amirhashchi 2011; Yadav and Saha 2012).

In this paper, we explore cosmological evolution for some well-known
$f(T)$ models by using BI universe. For this purpose, we evaluate
the EoS parameter and energy density of DE and display graphically.
The paper is organized as follows: In the next section, we review
teleparallel theory of gravity. We formulate the field equations of
$f(T)$ gravity and some cosmological parameters for BI universe in
section \textbf{3}. Section \textbf{4} is devoted to examine the
cosmological behavior of some $f(T)$ models to check whether the
crossing of phantom divide line occurs or not. We summarize and
conclude the results in the last section \textbf{5}.

\section{The $f(T)$ Formalism}

Here we briefly review the modified teleparallel theory of gravity.
In this theory, the vierbein field $\textbf{h}_i(x^\mu)$
at each point $x^{\mu}$ of the manifold are orthonormal basis,
where the Latin $(i,j,...=0,1,2,3)$ and Greek
alphabets $(\mu,\nu,...=0,1,2,3)$ are used for the tangent space
and spacetime indices respectively. The spacetime metric
tensor is defined from the dual vierbein as
\begin{equation}\label{1}
g_{\mu\nu}=\eta_{ij}h^{i}_{\mu}h^{j}_{\nu},
\end{equation}
where $\eta_{ij}=diag(1,-1,-1,-1)$ is the Minkowski spacetime for
the tangent space. For a given metric, there exists infinite tetrad
fields $h^{i}_{\mu}$ which satisfy the following properties (Ferraro
and Fiorini 2007; Hayashi and Shirafuji 1979)
\begin{equation}\label{2}
h^{i}_{\mu}h^{\mu}_{j}=\delta^{i}_{j}, \quad h^{i}_{\mu}h^{\nu}_{i}=\delta^{\nu}_{\mu}.
\end{equation}

The action for $f(T)$ theory with matter is given by (Bamba et al.
2011)
\begin{equation*}\label{3}
I=\frac{1}{16\pi G}\int d^4xe(T+f(T)+L_{m}).
\end{equation*}
Here $f(T)$ is a differentiable function of the torsion scalar
$T,~L_{m}$ is the matter Lagrangian and $e=\sqrt{-g}$. The torsion
scalar has the form
\begin{equation}\label{4}
T=S_{\rho}~^{\mu\nu}T^{\rho}~_{\mu\nu},
\end{equation}
here the torsion and the antisymmetric tensors are defined,
respectively by
\begin{eqnarray}\label{5}
T^{\rho}~_{\mu\nu}&=&\Gamma^{\rho}~_{\nu\mu}
-\Gamma^{\rho}~_{\mu\nu}=h^{\rho}_{i}
(\partial_{\mu}h^{i}_{\nu}-\partial_{\nu}h^{i}_{\mu}),\\
\label{6}S_{\rho}~^{\mu\nu}&=&\frac{1}{2}(K^{\mu\nu}~_{\rho}
+\delta^{\mu}_{\rho}T^{\theta\nu}~_{\theta}-\delta^{\nu}_{\rho}T^{\theta\mu}~_{\theta}),
\end{eqnarray}
where $\Gamma^{\rho}~_{\nu\mu}$ is the Weitzenb$\ddot{o}$ck
connection. The difference between Weitzenb$\ddot{o}$ck and
Levi-Civita connections is called contorsion tensor given by
\begin{equation}\label{7}
K^{\mu\nu}~_{\rho}=-\frac{1}{2}(T^{\mu\nu}~_{\rho}-T^{\nu\mu}~_{\rho}-T_\rho~^{\mu\nu}).
\end{equation}
The modified field equations of the teleparallel theory of gravity
(Sharif and Jawad 2011) are obtained by varying the action with
respect to the vierbien $h^{i}_{\mu}$ as (Bengochea and Ferraro
2009)
\begin{eqnarray}\label{8}
[e^{-1}\partial_{\mu}(eS_i~^{\mu\nu})-h_i^{\lambda}
T^{\rho}~_{\mu\lambda}S_{\rho}~^{\nu\mu}](1+f_T)+S_{i}~^{\mu\nu}
\partial_{u}(T)f_{TT}\nonumber\\
+\frac{1}{4}h^{\nu}_i(T+f(T))=\frac{1}{2}\kappa^2 h^{\rho}_i T_{\rho}^{\nu},
\end{eqnarray}
where $S_{i}~^{\mu\nu}=h_i^{\rho}
S_{\rho}~^{\mu\nu},~\kappa^{2}=8\pi G, ~f_{T}\equiv\frac{df}{dT}$.

\section{Bianchi I Universe and Some Cosmological Parameters}

The spatially homogenous and anisotropic LRS BI universe which has
one transverse direction $x$ and two equivalent longitudinal
directions $y$ and $z$, responsible for the anisotropic behavior is
(Sharif and Zubair 2010b)
\begin{equation}\label{11}
ds^2=dt^2-A^{2}(t) dx^2-B^{2}(t)(dy^2+ dz^2),
\end{equation}
where $A$ and $B$ are the cosmic scale factors. Using Eqs.(\ref{1}) and (\ref{11}),
the tetrad components are obtained as
\begin{equation}\label{12}
h_{\mu}^{i}=diag(1,A,B,B), \quad h_{i}^{\mu}=diag (1,A^{-1},
B^{-1},B^{-1}).
\end{equation}
Substituting Eqs.(\ref{5}) and (\ref{6}) in (\ref{4}), the torsion tensor for BI becomes
\begin{equation}\label{13}
T= -2\left(2\frac{\dot{A}\dot{B}}{AB}+ \frac{\dot{B^2}}{B^2}\right).
\end{equation}
The corresponding average scale factor $R$, the Hubble parameter $H$
and the anisotropy parameter $\Delta$ will become
\begin{eqnarray}\label{12}
R=(A B^2)^\frac{1}{3},\quad H=\frac{1}{3}\left(\frac{\dot{A}}{A}
+2\frac{\dot{B}}{B}\right),\quad
\Delta=\frac{1}{3}\sum_{i=1}^{3}\left(\frac{H_{i}-H}{H}\right)^{2},
\end{eqnarray}
where $H_{1}=\frac{\dot{A}}{A},~H_{2}=\frac{\dot{B}}{B}=H_{3}$ are
directional Hubble parameters along $x,~y$ and $z$-axes
respectively. The energy-momentum tensor $T_{\mu\nu}$ for perfect
fluid is
\begin{equation}\label{10}
T_{\rho}^{\nu}=diag(\rho_M,-P_M,-P_M,-P_M ),
\end{equation}
where $\rho_M$ and $P_M$ are the energy density and pressure of
matter inside the universe.

For $i=0=\nu$ and $i=1=\nu$ in Eq.(\ref{8}), we obtain the following
field equations
\begin{eqnarray}\label{14}
&&T+f(T)-4\left(2\frac{\dot{A}\dot{B}}{AB}+\frac{\dot{B^2}}{B^2}\right)(1+f_{T})
=2\kappa^{2}\rho_{M},\\\nonumber
&&4\left(\frac{\dot{A}\dot{B}}{AB}+\frac{\dot{B^2}}{B^2}+\frac{\ddot{B}}{B}\right)(1+f_{T})
-16\frac{\dot{B}}{B}\left[\frac{\dot{B}}{B}\left(\frac{\ddot{A}}{A}
-\frac{\dot{A^{2}}}{A^{2}}\right)\right.\\\label{15}
&&\left.+\left(\frac{\ddot{B}}{B}-\frac{\dot{B^{2}}}{B^{2}}\right)\left(\frac{\dot{B}}{B}
+\frac{\dot{A}}{A}\right)\right]f_{TT}-(T+f)=2\kappa^{2}P_{M}.
\end{eqnarray}
For a spatially homogeneous metric, the normal congruence to
homogeneous expansion implies that the expansion scalar $\Theta$ is
proportional to the shear scalar $\sigma$, i.e,
\begin{equation}\label{5*}
\Theta\varpropto\sigma.
\end{equation}
The expansion scalar and shear scalar for BI universe is given by
\begin{equation}\label{1*}
\Theta=\frac{\dot{A}}{A}+2\frac{\dot{B}}{B},\quad
\sigma=\frac{1}{\sqrt{3}}\left(\frac{\dot{A}}{A}-\frac{\dot{B}}{B}\right)
\end{equation}
This leads to the condition (Sharif and Waheed 2012; Bali and
Kumawat 2008; Amirhashchi 2011)
\begin{equation}\label{00+}
A(t)=B^m(t),\quad m\geq2.
\end{equation}
By using above condition, the anisotropy parameter of the expansion
is found to be
\begin{equation}\label{3++}
\Delta=2\frac{(m-1)^{2}}{(m+2)^{2}}.
\end{equation}
It is mentioned here that the isotropic behavior of the expanding
universe is obtained for $\Delta=0$. Under this supplementary
condition, the above field equations reduce to the following form
\begin{eqnarray}\label{16}
H^2&=&-\frac{(m+2)^2}{9(2m+1)}\left(\frac{8\pi G}{3}\rho_{M}
-\frac{f}{6}-\frac{Tf_{T}}{3}\right),
\\\label{17}(H^2)'&=&\frac{m+2}{3}\left(\frac{2\kappa^2 P_{M}
+\frac{2(m+2)}{2m+1}Tf_{T}+\frac{4m+5}{2m+1}T+f}
{2+2f_{T}+4Tf_{TT}}\right),
\end{eqnarray}
where prime denotes derivative with respect to
$\ln(B^\frac{m+2}{3})$. Notice that the above equations reduce to
the usual field equations for $T+f(T)=T$ i.e.
\begin{eqnarray}\label{18}
H^2&=&\frac{(m+2)^2}{9(2m+1)}\left[-\frac{8\pi G}{3}(\rho_{M}
+\rho_{DE})\right],\\\label{19}
(H^2)'&=&\left(\frac{m+2}{3}\right)8\pi G\left[P_{M}+P_{DE}
+\frac{4m+5}{3(2m+1)}(\rho_{M} +\rho_{DE})\right].
\end{eqnarray}

We assume only non-relativistic matter in which pressure is zero
i.e., $P_{M}=0$. Comparing Eqs.(\ref{16}) with (\ref{18}) and
(\ref{17}) with (\ref{19}), the energy density and pressure of the
effective DE become
\begin{eqnarray}\label{20}
\rho_{DE}&=&\frac{-1}{16\pi G}(f+2Tf_{T}),\\\label{21}
P_{DE}&=&\frac{1}{16\pi
G}\left(\frac{f-Tf_{T}-\frac{2(4m+5)}{2m+1}T^2f_{TT}}{1+f_{T}+2Tf_{TT}}\right).
\end{eqnarray}
Dividing Eq.(\ref{21}) by (\ref{20}), the EoS parameter for DE turns
out to be
\begin{equation}\label{22}
\omega_{DE}=-\frac{\frac{f}{T}-f_{T}-\frac{2(4m+5)}{2m+1}Tf_{TT}}{({1+f_{T}+2Tf_{TT}})
(\frac{f}{T}+2f_{T})}.
\end{equation}
The conservation laws of BI universe filled with pressureless matter and
DE take the following form
\begin{eqnarray*}\label{23}
\dot{\rho}_{M}+3H\rho_{M}=0,\\
\dot{\rho}_{DE}+3H\rho_{DE}(1+\omega_{DE})=0.
\end{eqnarray*}

\section{Cosmological Evolution in $f(T)$ Models}

In this section, we discuss the crossing of the phantom divide line
for the EoS parameter (\ref{22}) by using some particular $f(T)$
models. An appropriate scale factor should satisfy two requirements
from observations: one is the past transition from acceleration to
deceleration expansion phase and other is the currently slow
variation of the DE density. For the radiation, matter and DE epoch,
the universe can be characterized by a power law scale factor with
constant exponent. Due to the complexity of the field equations, it
is very difficult to evaluate the explicit analytical form of the
scale factor. So, we choose the scalar factor of the form (Sharif
and Rani 2011a; Radinschi 2000). to describe the possible supper
accelerated transition.
\begin{equation}\label{23*}
B(t) = (nst)^{1/n},
\end{equation}
where $n$ and $s$ are positive real constants.
For the sake of convenience, $s=1$.

\subsection{Model I}

Consider the exponential $f(T)$ model (Linder 2010; Jamil et al.
2011; Bamba et al. 2011)
\begin{equation}\label{24}
f(T)=\alpha T(1-e^{p T_{0}/T}),
\end{equation}
where
\begin{equation}\nonumber
\alpha=-\frac{1-\Omega^{(0)}_{M}}{1-(1-2p)e^p}, \quad p=constant.
\end{equation}
Here $T_{0}=T(z=0)$ is the current value of the torsion and the redshift is
\begin{equation}\label{27}
z \equiv 1/B^\frac{m+2}{3} -1.
\end{equation}
The current value of the fractional density $\Omega_{M}$ of
non-relativistic matter is defined by (Komatsu et al. 2011)
\begin{equation}\nonumber
\Omega^{(0)}_{m} \equiv \rho^{(0)}_{M}/\rho^{(0)}_{crit}=0.26,
\end{equation}
where $\rho^{(0)}_{M}$ is the current energy density and
$\rho^{(0)}_{crit}=3H^2_{0}/8\pi G$ is the critical density (Bamba
et al. 2010a) with current Hubble parameter $H_{0}=1$ (Sharif and
Jawad 2012). Notice that Eq.(\ref{24}) has only one parameter $p$ if
the value of $\Omega^{(0)}_{M}$ is known.

The EoS parameter in terms of $\mid T/T_{0} \mid$
can be written as
\begin{equation}\label{1003}
\omega_{DE}=\frac{-e^{pT_{0}/T}}{E}\left(\frac{pT_{0}}{T}\right)
\left[-1+\frac{2(4m+5)}{2m+1}\left(\frac{pT_{0}}{T}\right)\right],
\end{equation}
where
\begin{eqnarray}\nonumber
E&=&\left[1+\alpha (1-e^{pT_{0}/T})+\alpha (pT_{0}/T)e^{pT_{0}/T}(1-2pT_{0}/T)\right]\nonumber\\
&\times&\left[3(1-e^{pT_{0}/T})+2(pT_{0}/T)e^{pT_{0}/T}\right]\nonumber.
\end{eqnarray}
\begin{figure}
\epsfig{file=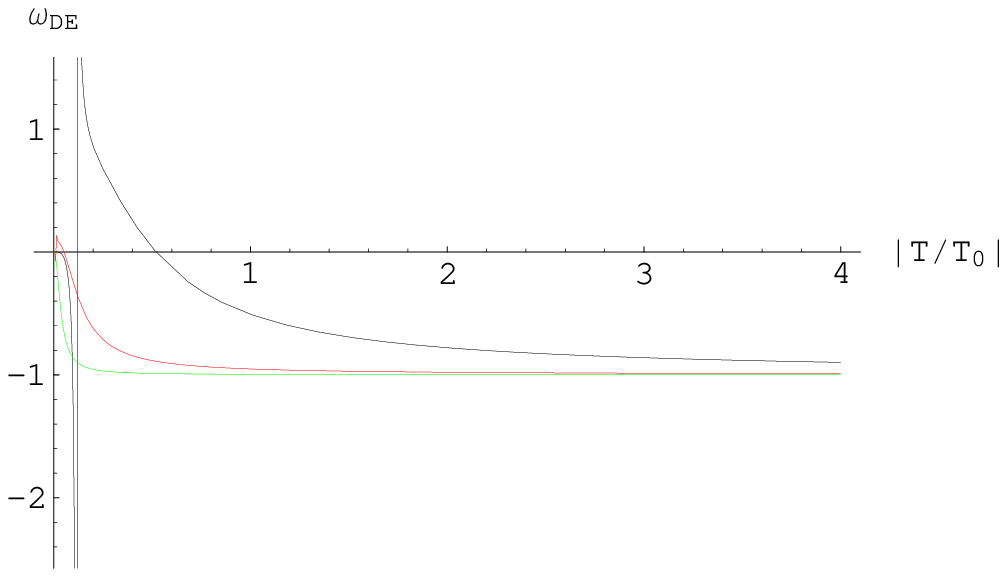, width=0.5\linewidth}\epsfig{file=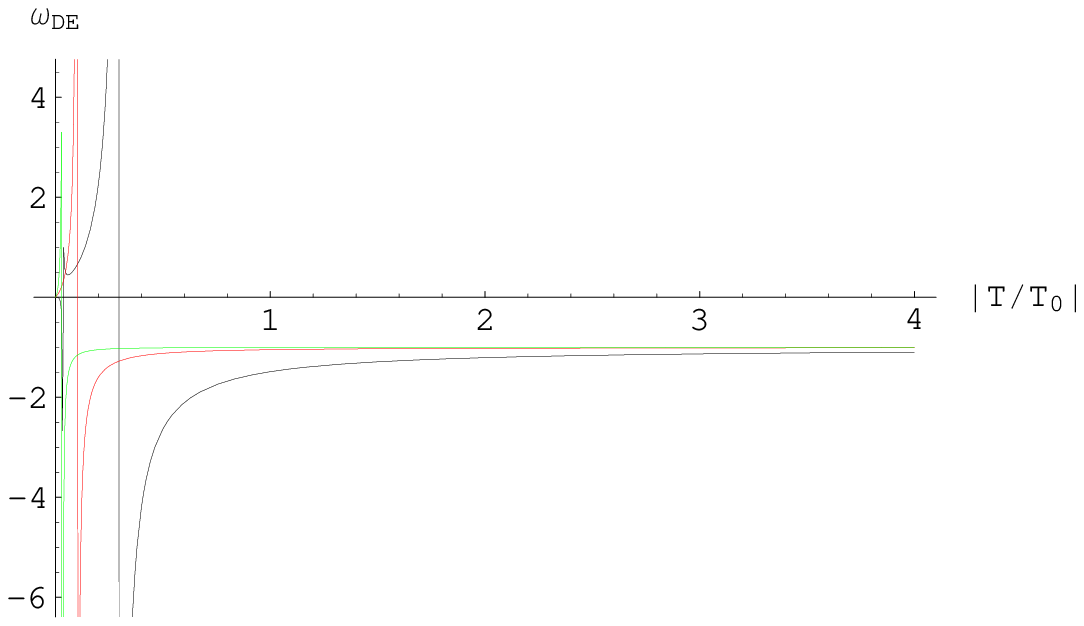,
width=0.5\linewidth}\caption{Plot of $\omega_{DE}$ versus $\mid
T/T_{0} \mid$ with $m=2$ for model I. In left graph, $|p|=0.1$
(black), $0.01$ (red), $0.001$ (green) and in right graph $|p|=-0.1$
(black), $-0.01$ (red), $-0.001$ (green).}
\end{figure}
The graphical behavior of $\omega_{DE}$ as a function of $\mid
T/T_{0}\mid$ is shown in \textbf{Figure 1}. For both $p>0$ and
$p<0,~\omega_{DE}$ reaches to $-1$ but does not cross the phantom
divide line $(\omega_{DE}=-1)$. Thus the universe stays in DE era as
$\mid T/T_{0}\mid$ approaches to infinity.

Inserting Eq.(\ref{00+}) in (\ref{13}) and using (\ref{23*}) and (\ref{27}),
we get a torsion scalar as a function of redshift $z$
\begin{equation}\label{2*}
T=-2(2m+1)(1+z)^\frac{6n}{m+2}, \quad T_{0}=-2(2m+1).
\end{equation}
Substituting these values in Eq.(\ref{1003}), the corresponding EoS
parameter in terms of $z$ takes the following form as
\begin{equation}
\omega_{DE}=-\frac{pe^{p(1+z)^{\frac{-6n}{m+2}}}}{D(1+z)^{\frac{6n}{m+2}}}
\left[-1+\frac{2(4m+5)p} {(2m+1)(1+z)^{\frac{6n}{m+2}}}\right],
\end{equation}
where
\begin{eqnarray}\nonumber
D&=&\left[3\left(1-e^{p(1+z)^{\frac{-6n}{m+2}}}\right)
+2p(1+z)^{\frac{-6n}{m+2}}e^{p(1+z)^{\frac{-6n}{m+2}}}\right]\nonumber\\
&\times&\left[1+\alpha\left(1-e^{p(1+z)^{\frac{-6n}{m+2}}}\right)+\alpha
p(1+z)^{\frac{-6n}{m+2}}e^{p(1+z)^{\frac{-6n}{m+2}}}\right.\nonumber\\
&\times&\left.\left(1-2p(1+z)^{\frac{-6n}{m+2}}\right)\right].\nonumber
\end{eqnarray}
\begin{figure}
\epsfig{file=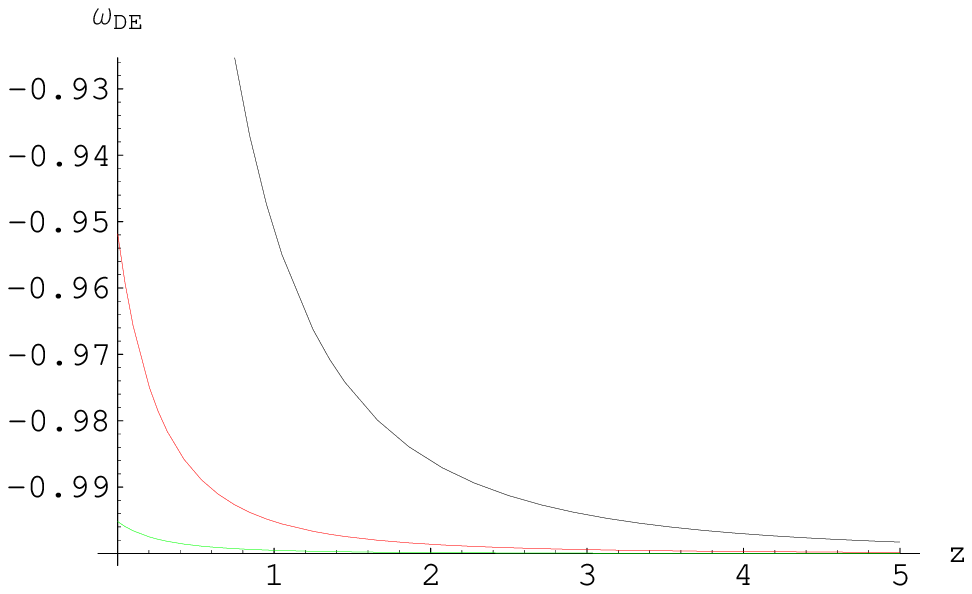, width=0.5\linewidth} \epsfig{file=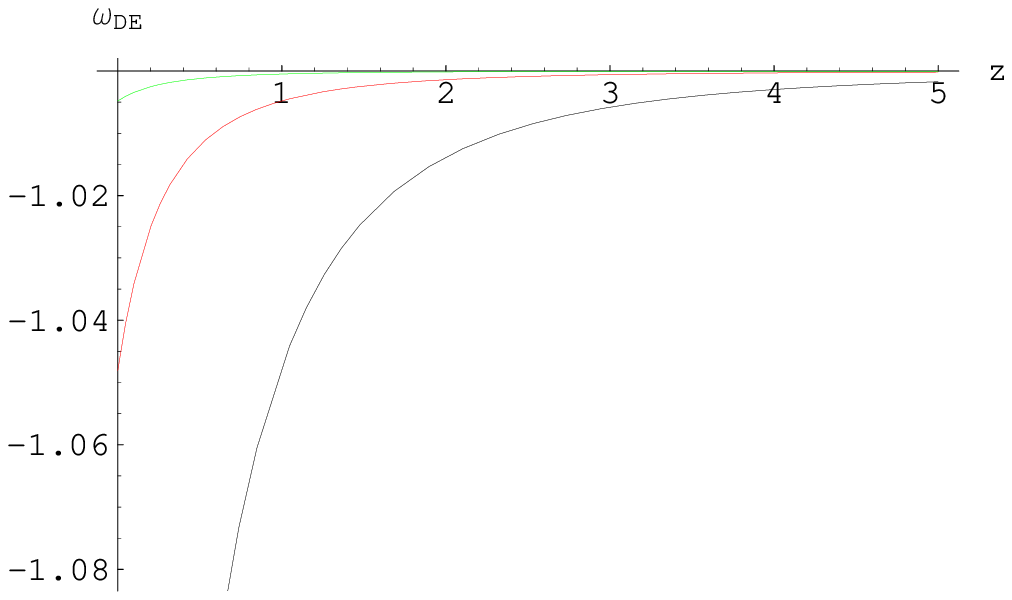,
width=0.5\linewidth} \caption{Plot of $\omega_{DE}$ versus redshift
$z$ for model I with $m=2,~n=2$ and the same values of the parameter
$p$ as in Figure 1.}
\end{figure}
The graphical representation of $\omega_{DE}$ versus redshift is
shown in \textbf{Figure 2}. We see from the left graph that for
$p>0,~\omega_{DE}$ increases towards negative and attains $-1$ but
does not cross the phantom divide line for $z\rightarrow\infty$.
Thus the universe stays in non-phantom phase (quintessence). For
$p<0,~\omega_{DE}$ is less than $-1$ without crossing the phantom
divide line and gives the phantom phase of the universe initially.
As $z\rightarrow\infty$, it reaches to $-1$ which is shown in the
right graph of \textbf{Figure 2}. Thus the universe always stays in
DE era for both cases.

The parameter $\rho_{DE}^{(*)}\equiv \rho_{DE}/\rho_{DE}^{(0)}$ in
terms of $z$ is obtained by putting Eq.(\ref{2*}) in (\ref{20}) as
\begin{equation}\label{20+}
\rho_{DE}^{(*)}=\frac{(2m+1)(1+z)^\frac{6n}{m+2}\alpha}{8\pi G \rho_{DE}^{(0)}}
\left[3\left(1-e^{p/(1+z)^{\frac{6n}{m+2}}}
\right)+\frac{2pe^{p/(1+z)^{\frac{6n}{m+2}}}}{(1+z)^{\frac{6n}{m+2}}}\right],
\end{equation}
where $\rho_{DE}^{(0)}=0.74~\rho^{(0)}_{crit}$ (Bamba et al. 2011).
\begin{figure}
\epsfig{file=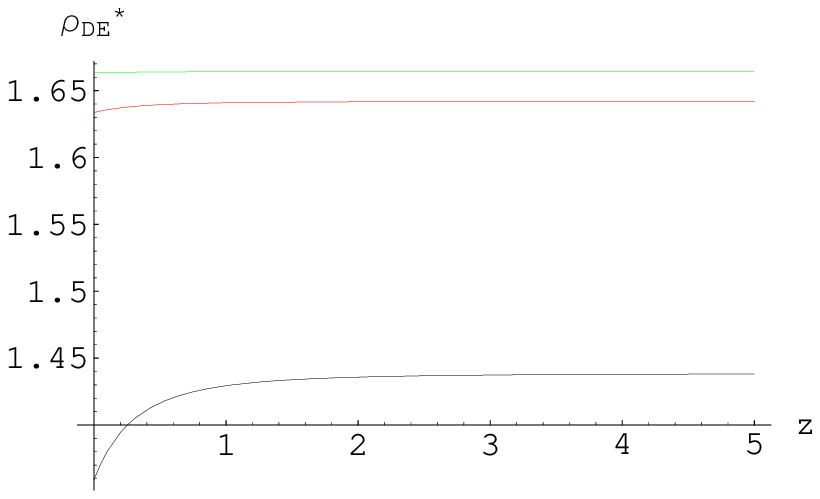, width=0.5\linewidth}\epsfig{file=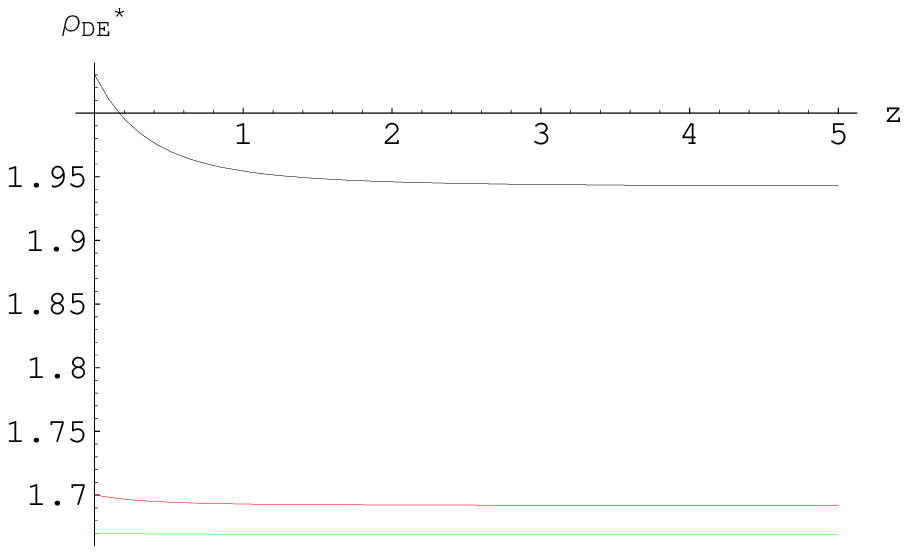,
width=0.5\linewidth}\caption{Plot of $\rho_{DE}^{(*)}$ versus $z$
keeping same parameters as in Figure 2.}
\end{figure}
The evolution of $\rho_{DE}^{(*)}$ in terms of redshift $z$ for
$p>0$ and $p<0$ is shown in \textbf{Figure 3}. For $p>0$, this
indicates a slight increment in $\rho_{DE}^{(*)}$ for smaller values
of $z$ and becomes constant for larger values of $z$. When $p<0$, it
decreases initially with respect to $z$ and approaches to constant
value as $z\rightarrow \infty$. We would like to mention here that
$\rho_{DE}^{(*)}$ attains different values at $z=0$ for both the
cases of $p$ for BI universe. On the other hand,
$\rho_{DE}^{(*)}(z=0)$ coincides at the same point for several
values of $p$ for FRW universe (Bamba et al. 2011).

Now we discuss the viability of the exponential $f(T)$ model for
phantom and non-phantom phases by using an approximate method. From
Eq.(\ref{24}), we obtain
\begin{equation}\label{20*}
f_{T}=\alpha\left(1-e^{pT_{0}/T}+\frac{pT_{0}}{T}e^{pT_{0}/T}\right), \quad
f_{TT}=-\alpha\left(\frac{pT_{0}}{T}\right)^{2}\frac{1}{T}e^{pT_{0}/T}.
\end{equation}
Assuming $X=pT_{0}/T,~T_{0}/T\lesssim1$ in Eqs.(\ref{24}) and
(\ref{20*}), it follows that (Bamba et al. 2010b)
\begin{equation}\label{22*}
\frac{f}{T}\approx-\alpha\left(X+\frac{X^{2}}{2}\right), \quad f_{T}\approx\frac{\alpha X^{2}}{2},
\quad Tf_{TT}\approx-\alpha X^{2}.
\end{equation}
Inserting these values in Eq.({\ref{22}), $\omega_{DE}$ takes the
form
\begin{equation}\label{25*}
\omega_{DE}\approx-1+\frac{14m+19}{2(2m+1)}X.
\end{equation}
Here, we have take $\alpha\sim O(1)$. This implies that the behavior
of EoS parameter depends on the sign of $p$ and correspondingly on
$X$. For $X>0$, the universe always stays in non-phantom phase as
$\omega_{DE}>-1$. The universe rests in the  phantom region for
$X<0$ as $\omega_{DE}<-1$. This is consistent with the graphical
results shown in \textbf{Figure 2}.

\subsection{Model II}

Assume the logarithmic $f(T)$ model as (Bamba et al. 2011)
\begin{equation}\label{28}
f(T)=\beta T_{0}\left(\frac{qT_{0}}{T}\right)^{-1/2}\ln\left(\frac{qT_{0}}{T}\right),
\end{equation}
where
\begin{equation}\nonumber
\beta \equiv \frac{1-\Omega^{(0)}_{M}}{2q^{-1/2}}, \quad q>0.
\end{equation}
If the value of $\Omega^{(0)}_{M}$ is given, then the only parameter
$q$ is involved in the logarithmic $f(T)$ model same as the
exponential $f(T)$ model. The corresponding EoS parameter is given
by
\begin{equation}\label{30}
\omega_{DE}=-\frac{[1+\frac{3(m+1)}{2m+1}\ln(T_{0}/T)]}{[2-(1-\Omega^{(0)}_{M})
(T_{0}/T)^{1/2}][\ln(T_{0}/T) -1]}
\end{equation}
which is independent of $q$. Using Eq.(\ref{2*}) in above equation,
$\omega_{DE}$ in terms of redshift $z$ becomes
\begin{equation}\label{30+}
\omega_{DE}=-\frac{[1+\frac{3(m+1)}{2m+1}\ln(1/(1+z)^{\frac{6n}{m+2}})]}{[2-(1-\Omega^{(0)}_{M})
(1/(1+z)^{\frac{6n}{m+2}})^{1/2}][\ln(1/(1+z)^{\frac{6n}{m+2}})-1]}.
\end{equation}
\begin{figure}
\epsfig{file=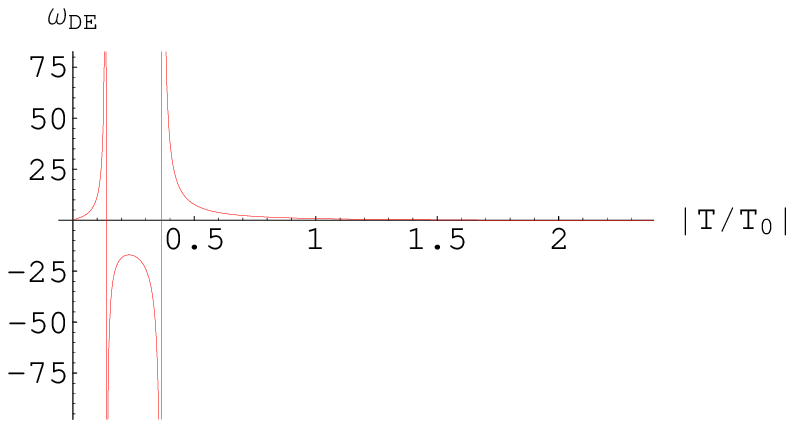, width=0.5\linewidth}\epsfig{file=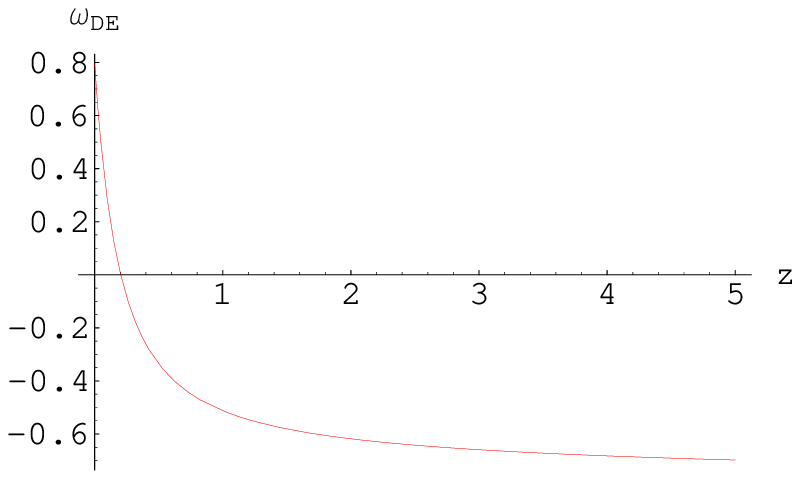,
width=0.5\linewidth}\caption{Plot of $\omega_{DE}$ versus $\mid
T/T_{0}\mid$ in left side and $\omega_{DE}$ versus $z$ in right side
for $q=1,~m=2=n$ and $\Omega^{(0)}_{M} = 0.26$ for model II.}
\end{figure}
Its behavior is shown in \textbf{Figure 4}. The left graph shows
that $\omega_{DE}$ becomes negative as $\mid
T/T_{0}\mid\rightarrow\infty$. In the beginning, the model
represents a universe having both properties of matter and
radiation. After a very short interval, the universe enters in DE
phase. It is mentioned here that the universe remains in non-phantom
phase as the time elapses. The right graph has the same behavior as
that of the exponential $f(T)$ model and the universe stays in the
non-phantom phase ($\omega_{DE}>-1$).

\subsection{Model III}

Here we take the combination of both exponential and logarithmic
$f(T)$ models which has the following form (Bamba et al. 2011)
\begin{equation}\label{32+}
f(T)=\gamma\left[T_{0}\left(\frac{uT_{0}}{T}\right)^{-1/2}
\ln\left(\frac{uT_{0}}{T}\right)-T(1-e^{uT_{0}/T})\right],
\end{equation}
where
\begin{equation}\nonumber
\gamma \equiv \frac{1-\Omega^{(0)}_{M}}{2u^{-1/2}+[1-(1-2u)e^{u}]}.
\end{equation}
The positive constant $u$ is the only parameter in this model.
The EoS parameter for DE in terms of $\mid T/T_{0} \mid$ is given by
\begin{eqnarray}\label{4*}
\omega_{DE}&=&-\frac{1}{I}\left[\frac{1}{u}\sqrt\frac{uT_{0}}{T}
\left\{1+\frac{3(m+1)}{2m+1}\ln\left(\frac{uT_{0}}{T}\right)\right\}\right.\nonumber\\
&+&\left. e^{uT_{0}/T}\left(\frac{uT_{0}}{T}\right)\left\{1-
\frac{2(4m+5)}{2m+1}\left(\frac{uT_{0}}{T}\right)\right\}\right],
\end{eqnarray}
where
\begin{eqnarray}\nonumber
I&=&\left[\frac{2}{u}\sqrt\frac{uT_{0}}{T}
\left\{\ln\left(\frac{uT_{0}}{T}\right)-1\right\}+e^{uT_{0}/T}
\left(3-\frac{2uT_{0}}{T}\right)-3\right]\nonumber\\
&\times&\left[1-\gamma(1-e^{uT_{0}/T})
-\gamma\sqrt\frac{uT_{0}}{T}\left\{\frac{1}{u}+\sqrt\frac{uT_{0}}{T}e^{uT_{0}/T}\left(1-
\frac{2uT_{0}}{T}\right)\right\}\right].\nonumber
\end{eqnarray}
\begin{figure}
\epsfig{file=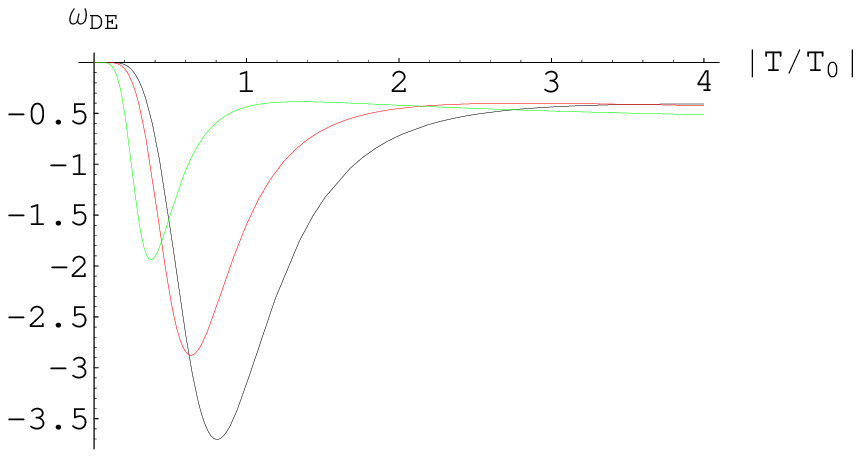, width=0.5\linewidth} \epsfig{file=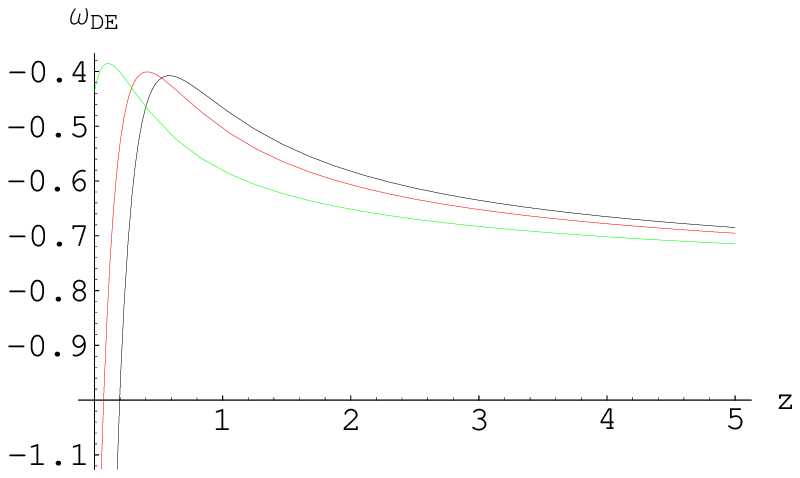,
width=0.5\linewidth}\caption{Plot of $\omega_{DE}$ versus $\mid
T/T_{0} \mid$ in left graph and $\omega_{DE}$ versus $z$ in right
graph with $m=2=n$ and $u=1$ (black line), $0.8$ (red line), $0.5$
(green) for model III.}
\end{figure}
Initially, the expanding universe is lying in non-phantom phase with
$\omega_{DE}$ as a function of $\mid T/T_{0} \mid$ as shown in left
graph of \textbf{Figure 5}. The EoS parameter decreases with
increase in $\mid T/T_{0} \mid$ by crossing the phantom divide line
at $\mid T/T_{0} \mid=0.27,~0.36$ and $0.42$ for $u=0.5,~0.8$ and
$1$ respectively evolving the phantom phase. After a short interval,
$\omega_{DE}$ crosses the phantom divide line again at $\mid
T/T_{0}\mid=0.65,~1.2$ and $1.7$ which turns out to be constant as
$\mid T/T_{0} \mid$ increases.

Inserting Eq.(\ref{2*}) in (\ref{4*}), $\omega_{DE}$ in terms of $z$
turns to be
\begin{eqnarray}\nonumber
\omega_{DE}&=&-\frac{1}{F}\left[\frac{1}{u}{\sqrt{u(1+z)^{\frac{-6n}{m+2}}}}
\left\{1+\frac{3(m+1)}{2m+1}\ln\left(u(1+z)^{\frac{-6n}{m+2}}
\right)\right\}\right.\nonumber\\
&+&\left.u(1+z)^{\frac{-6n}{m+2}}e^{u(1+z)^{\frac{-6n}{m+2}}}\left(1-\frac{2u(4m+5)(1+z)^{\frac{-6n}
{m+2}}}{(2m+1)}\right)\right],
\end{eqnarray}
where
\begin{eqnarray}\nonumber
F&=&\left[\frac{2}{u}{\sqrt{u(1+z)^{\frac{-6n}{m+2}}}}
\left(\ln\left(u(1+z)^{\frac{-6n}{m+2}}\right)-1\right)
+e^{u(1+z)^{\frac{-6n}{m+2}}}\right.\nonumber\\
&\times&\left.\left\{3-2u(1+z)^{\frac{-6n}{m+2}}\right\}-3\right]
\left[1-\gamma\left(1-e^{u(1+z)^{\frac{-6n}{m+2}}}\right)
-\gamma{\sqrt{u(1+z)^{\frac{-6n}{m+2}}}}\right.\nonumber\\
&\times&\left.\left\{\frac{1}{u}+e^{u(1+z)^{\frac{-6n}{m+2}}}\left(1-2u(1+z)^{\frac{-6n}{m+2}}\right)
{\sqrt{u(1+z)^{\frac{-6n}{m+2}}}}\right\}\right].\nonumber
\end{eqnarray}
The right graph of \textbf{Figure 5} shows that for $u=0.8$ and
$u=1$, the universe is in the phantom phase ($\omega_{DE}<-1$) at
initial epoch. As $z$ increases, it crosses the phantom divide line
($\omega_{DE}=-1$) at $z=0.067$ and $0.18$ respectively. Thus
$\omega_{DE}$ enters in the non-phantom phase and converges to
constant value with increment in $z$. It is interesting to note that
for $u \leq 0.6$, the universe always rests in non-phantom phase.
Notice that the combined $f(T)$ model behaves as the quintom model
(Khatua et al. 2011).

\subsection{Model IV}

Now we take the $f(T)$ model of the type (Yang 2011)
\begin{equation}\label{35}
f(T)=T+\eta T_{0}\frac{(T^{2}/T_{0}^{2})^{\lambda}}{1+(T^{2}/T_{0}^{2})^{\lambda}},
\end{equation}
where $\eta$ and $\lambda$ are positive constants. The corresponding
EoS parameter is
\begin{eqnarray}\nonumber
\omega_{DE}&=&-\frac{\eta
(T/T_{0})^{2\lambda-1}}{K(1+(T^{2}/T_{0}^{2})^{\lambda})}
\left[1-2\lambda\left\{1-\frac{(T^{2}/T_{0}^{2})^{\lambda}}
{(1+(T^{2}/T_{0}^{2})^{\lambda})}\right.\right.\nonumber\\
&+&\left.\left.\left(2\lambda-1-\frac{(T^{2}/T_{0}^{2})^{\lambda}}
{1+(T^{2}/T_{0}^{2})^{\lambda}}\left(6\lambda-1-\frac{4\lambda(T^{2}/T_{0}^{2})^{\lambda}}
{1+(T^{2}/T_{0}^{2})^{\lambda}}\right)\right)\right.\right.\nonumber\\
&\times&\left.\left.\frac{2(4m+5)}{2m+1}\right\}\right],\label{26}
\end{eqnarray}
where
\begin{eqnarray}\nonumber
K&=&\left[3+\frac{\eta(T/T_{0})^{2\lambda-1}}{1+(T^{2}/T_{0}^{2})^{\lambda}}
\left\{1+4\lambda\left(1-\frac{(T^{2}/T_{0}^{2})^{\lambda}}
{1+(T^{2}/T_{0}^{2})^{\lambda}}\right)\right\}\right]\nonumber\\
&\times&\left[1+\frac{2\eta\lambda(T/T_{0})^{2\lambda-1}}
{1+(T^{2}/T_{0}^{2})^{\lambda}}
\left\{4\lambda-1+\frac{(T^{2}/T_{0}^{2})^{\lambda}}
{1+(T^{2}/T_{0}^{2})^{\lambda}}\right.\right.\nonumber\\
&\times&\left.\left.
\left(12\lambda-1+\frac{8\lambda(T^{2}/T_{0}^{2})^{\lambda}}
{1+(T^{2}/T_{0}^{2})^{\lambda}}\right)\right\}\right].\nonumber
\end{eqnarray}
\begin{figure}
\epsfig{file=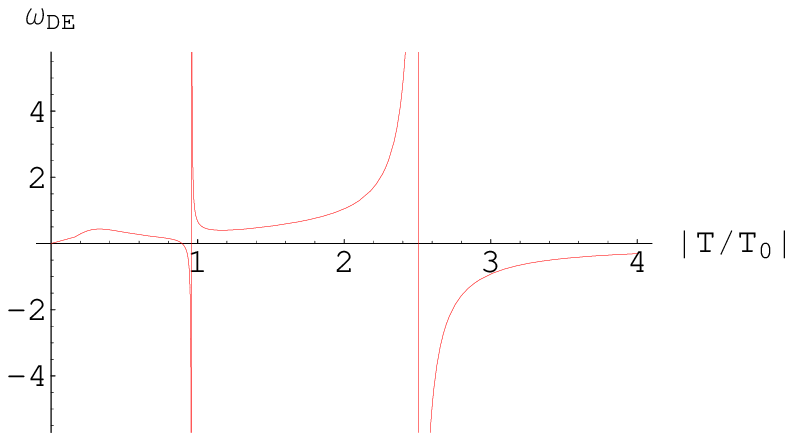, width=0.5\linewidth} \epsfig{file=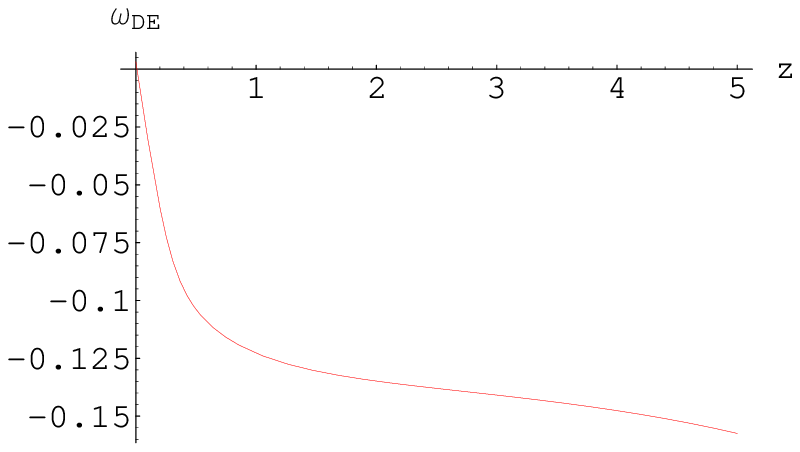,
width=0.5\linewidth}\caption{Plot of $\omega_{DE}$ versus $\mid
T/T_{0} \mid$ and $\omega_{DE}$ versus $z$ for $m=2=n,~\eta=3$ and
$\lambda=2$ for model IV.}
\end{figure}
The left graph of $\omega_{DE}$  as a function of $\mid T/T_{0}
\mid$ is shown in \textbf{Figure 6}. Initially, the behavior of
$\omega_{DE}$ is positive and stays in radiation era for very small
region. After a slight variation in $\mid T/T_{0} \mid$, the
crossing of phantom divide line occurs and results a phantom era.
There are two singularities at $\mid T/T_{0}\mid=0.95$ and $2.5$
with positive behavior. After these singularities, $\omega_{DE}$
evolves from phantom to non-phantom phase by crossing phantom divide
line and remains in non-phantom phase for $\mid
T/T_{0}\mid\rightarrow\infty$ causing accelerated expansion of the
universe. In term of $z$, the parameter $\omega_{DE}$ is obtained by
using Eq.(\ref{2*}) in (\ref{26}) as
\begin{eqnarray}\nonumber
\omega_{DE}&=&-\frac{\eta
(1+z)^{\frac{6n(2\lambda-1)}{m+2}}}{J(1+(1+z)^{\frac{12n\lambda}{m+2}})}
\left[1-2\lambda\left\{1-\frac{(1+z)^{\frac{12n\lambda}{m+2}}}
{1+(1+z)^{\frac{12n\lambda}{m+2}}}+\frac{2(4m+5)}{2m+1}\right.\right.\nonumber\\
&\times&\left.\left.\left(2\lambda-1-\frac{(1+z)^{\frac{12n\lambda}{m+2}}}
{1+(1+z)^{\frac{12n\lambda}{m+2}}}\left(6\lambda-1-4\lambda
\frac{(1+z)^{\frac{12n\lambda}{m+2}}}
{1+(1+z)^{\frac{12n\lambda}{m+2}}}\right)\right)\right\}\right],\nonumber\\
\end{eqnarray}
where
\begin{eqnarray}\nonumber
J&=&\left[3+\frac{\eta(1+z)^{\frac{6n(2\lambda-1)}{m+2}}}{1+(1+z)^{\frac{12n\lambda}{m+2}}}
\left\{1+4\lambda\left(1-\frac{(1+z)^{\frac{12n\lambda}{m+2}}}
{1+(1+z)^{\frac{12n\lambda}{m+2}}}\right)\right\}\right]\nonumber\\
&\times&\left[1+\frac{2\eta\lambda(1+z)^{\frac{6n(2\lambda-1)}{m+2}}}
{1+(1+z)^{\frac{12n\lambda}{m+2}}}
\left\{4\lambda-1+\frac{(1+z)^{\frac{12n\lambda}{m+2}}}
{1+(1+z)^{\frac{12n\lambda}{m+2}}}\right.\right.\nonumber\\
&\times&\left.\left.\left(12\lambda-1+\frac{8\lambda(1+z)^{\frac{12n\lambda}{m+2}}}
{1+(1+z)^{\frac{12n\lambda}{m+2}}}\right)\right\}\right].\nonumber
\end{eqnarray}
For small values of $z$, $\omega_{DE}$ is found to be the
non-phantom phase and converges to zero with the increment in the
value of $z$. This shows that matter becomes dominant over the DE as
shown in right side of \textbf{Figure 6}. Inserting Eq.(\ref{2*}) in
(\ref{20}), the corresponding parameter $\rho_{DE}^{(*)}(z)$ can be
written as
\begin{eqnarray}\label{20+}
\rho_{DE}^{(*)}&=&\frac{(2m+1)(1+z)^\frac{6n}{m+2}}{8\pi G \rho_{DE}^{(0)}}
\left[3+\frac{\eta (1+z)^{\frac{6n(2\lambda-1)}{m+2}}}
{1+(1+z)^{\frac{12n\lambda}{m+2}}}\right.\nonumber\\
&\times&\left.\left\{1+4\lambda\left(1-\frac{(1+z)^{\frac{12n\lambda}{m+2}}}
{1+(1+z)^{\frac{12n\lambda}{m+2}}}\right)\right\}\right].
\end{eqnarray}
\begin{figure}
\center\epsfig{file=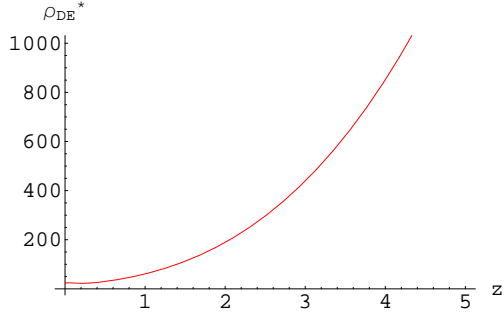, width=0.5\linewidth}\caption{Plot of
$\rho_{DE}^{(*)}$ versus $z$ with same values of parameters as in
Figure 6.}
\end{figure}
The graphical representation (\textbf{Figure 7}) shows the
increasing behavior of $\rho_{DE}^{(*)}$ for $z\rightarrow\infty$.

\subsection{Model V}

Finally, we take the $f(T)$ model in the form (Yang 2011)
\begin{equation}\label{36}
f(T)=T-\eta
T_{0}\left[\left(1+\frac{T^{2}}{T_{0}^{2}}\right)^{-\lambda}-1\right].
\end{equation}
The expression for $\omega_{DE}$ takes the form
\begin{eqnarray}\nonumber
\omega_{DE}&=&\frac{1}{L}\left[\eta\left(\frac{T}{T_{0}}\right)
\left\{1-\left(1+\frac{T^{2}}{T^{2}_{0}}\right)^{-\lambda}\right\}
-2\eta\lambda\left(\frac{T}{T_{0}}\right)\left(1+\frac{T^{2}}{T^{2}_{0}}
\right)^{-\lambda-1}\right.\nonumber\\
&\times&\left.
\left\{1+\frac{2(4m+5)}{2m+1}\left(1-2(\lambda+1)\left(\frac{T}{T_{0}}\right)^{2}
\left(1+\frac{T^{2}}{T^{2}_{0}}\right)^{-1}\right)\right\}\right],\label{27+}
\end{eqnarray}
where
\begin{eqnarray}\nonumber
L&=&\left[2+2 \eta \lambda \left(\frac{T}{T_{0}}\right)
\left(1+\frac{T^{2}}{T^{2}_{0}}\right)^{-\lambda-1}
\left\{3-4(\lambda+1)\left(\frac{T}{T_{0}}\right)^{2}
\left(1+\frac{T^{2}}{T^{2}_{0}}\right)^{-1}\right\}\right]\nonumber\\
&\times&\left[3-\eta\left(\frac{T_{0}}{T}\right)
\left(1+\frac{T^{2}}{T^{2}_{0}}\right)^{-\lambda}
\left\{1-4\lambda\left(\frac{T}{T_{0}}\right)^{2}
\left(1+\frac{T^{2}}{T^{2}_{0}}\right)^{-1}\right\}+\eta\left(\frac{T_{0}}{T}
\right)\right].\nonumber
\end{eqnarray}
The left graph of \textbf{Figure 8} represents the cosmological
evolution of $\omega_{DE}$ in terms of $\mid T/T_{0}\mid$ which
shows the same behavior as \textbf{model IV}. However, the
singularities appear at $\mid T/T_{0}\mid=0.7,~1.7$ and the universe
stays in non-phantom phase for a short interval. Here the universe
becomes matter dominated for higher values of $\mid T/T_{0} \mid$.

Substituting Eq.(\ref{2*}) in (\ref{27+}), the corresponding
$\omega_{DE}$ as a function of $z$ turns out to be
\begin{eqnarray}\nonumber
\omega_{DE}&=&\frac{1}{M}\left[-2\eta\lambda(1+z)^\frac{6n}{m+2}
\left(1+(1+z)^{\frac{12n}{m+2}}\right)^{-\lambda-1}\right.\nonumber\\
&\times&\left.\left\{1+\frac{2(4m+5)}{2m+1}
\left(1-2(\lambda+1)(1+z)^\frac{12n}{m+2}
\left(1+(1+z)^{\frac{12n}{m+2}}\right)^{-1}\right)\right\}\right.\nonumber
\\&+&\left.\eta\left((1+z)^{\frac{6n}{m+2}}\right)
\left\{1-\left(1+(1+z)^{\frac{12n}{m+2}}\right)^{-\lambda}\right\}\right],
\end{eqnarray}
where
\begin{eqnarray}\nonumber
M&=&\left[1+\eta\lambda(1+z)^\frac{6n}{m+2}
\left(1+(1+z)^\frac{12n}{m+2}\right)^{-\lambda-1}\right.\nonumber\\
&\times&\left.\left\{3-4(\lambda+1)(1+z)^\frac{12n}{m+2}
\left(1+(1+z)^\frac{12n}{m+2}\right)^{-1}\right\}\right]\nonumber\\
&\times&\left[3-\frac{\eta}{(1+z)^{\frac{6n}{m+2}}}
\left(1+(1+z)^{\frac{12n}{m+2}}\right)^{-\lambda}\right.\nonumber\\
&\times&\left.\left\{1-4\lambda(1+z)^{\frac{12n}{m+2}}
\left(1+(1+z)^{\frac{12n}{m+2}}\right)^{-1}\right\}+\frac{\eta}{(1+z)^{\frac{6n}{m+2}}}
\right].\nonumber
\end{eqnarray}
\begin{figure}
\epsfig{file=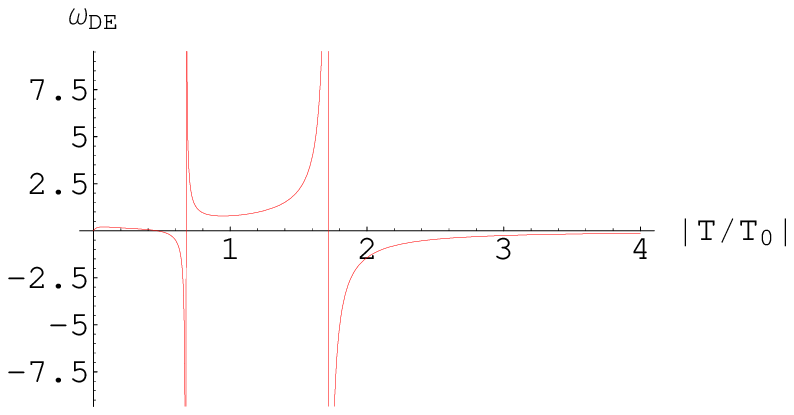, width=0.5\linewidth} \epsfig{file=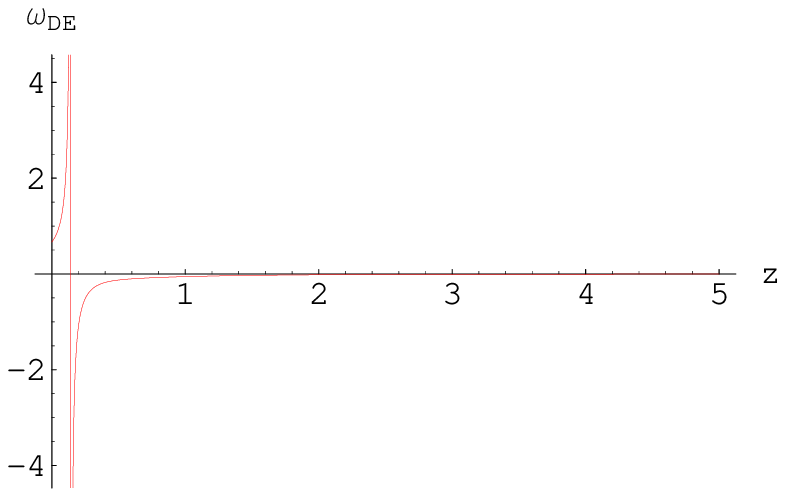,
width=0.5\linewidth}\caption{Plot of $\omega_{DE}$ versus $\mid
T/T_{0}\mid$ in the left and $\omega_{DE}$ versus $z$ in the right
for $\eta=3,~\lambda=3$ and $m=2=n$ for model V.}
\end{figure}
The parameter $\omega_{DE}$ in terms of $z$ enters from phantom
phase to non-phantom phase and approaches to matter dominated era
similar to $\omega_{DE}(T/T_{0})$ shown in the right graph of
\textbf{Figure 8}. Using Eq.(\ref{2*}) in (\ref{20}), it follows
that
\begin{eqnarray}\nonumber
\rho_{DE}^{(*)}(z)&=&\frac{(2m+1)(1+z)^\frac{6n}{m+2}}{8\pi G \rho_{DE}^{(0)}}
\left[3-\frac{\eta}{(1+z)^{\frac{6n}{m+2}}}
\left(1+(1+z)^{\frac{12n}{m+2}}\right)^{-\lambda}\right.\nonumber\\
&\times&\left.\left\{1-4\lambda(1+z)^{\frac{12n}{m+2}}
\left(1+(1+z)^{\frac{12n}{m+2}}\right)^{-1}\right\}+\frac{\eta}{(1+z)^{\frac{6n}{m+2}}}\right].
\end{eqnarray}
\begin{figure}
\center\epsfig{file=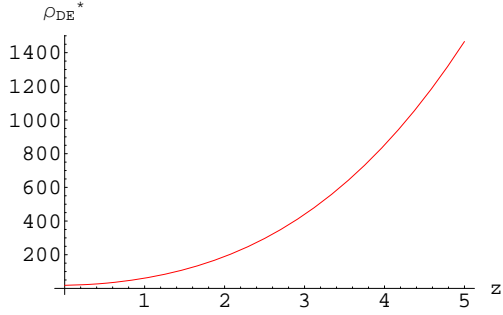, width=0.5\linewidth}\caption{Plot of
$\rho_{DE}^{(*)}$ versus $z$ for $\eta=3$ and $\lambda=2$}
\end{figure}
\textbf{Figure 9} shows that $\rho_{DE}^{(*)}$ is
positive and increases with $z\rightarrow\infty$.

\section{Summary and Conclusion}

The main purpose of this paper is to discuss the well-known
phenomenon of the universe expansion in the context of $f(T)$
gravity for BI universe model. We have investigated the cosmological
evolution of EoS parameter $\omega_{DE}$ and energy density
$\rho_{DE}^{(*)}$ for some well-known $f(T)$ models. These
parameters are evaluated in terms of $\mid T/T_{0}\mid$ and redshift
$z$. The graphical representation of the phantom and non-phantom
phases of the expanding universe is examined. The behavior of these
models can be summarized as follows:
\begin{itemize}
\item  Model I is the exponential $f(T)$ model in which the phase
of the universe changes with the sign of parameter $p$. For
$p>0,~\omega_{DE}(z)$ indicates a universe dominated by the
non-phantom era while the phantom phase is obtained for $p<0$. The
EoS parameter $\omega_{DE}(\mid T/T_{0} \mid)$ corresponds to the DE
phase. The energy density shows positive behavior both for $p>0$ and
$p<0$.
\item  For the logarithmic model II, $\omega_{DE}$ as a function of $\mid T/T_{0}\mid$
and $z$ remains in non-phantom phase which shows the accelerated
expansion of the universe.
\item In the combined model III, $\omega_{DE}(\mid T/T_{0}\mid)$ shows that
the universe initially transits from the non-phantom to phantom
phase. After that, the universe again crosses the phantom divide
line and retains its initial phase. On the other hand,
$\omega_{DE}(z)$ crosses the phantom divide line from phantom phase
to non-phantom phase only once.
\item For model IV, the parameter $\omega_{DE}(\mid
T/T_{0} \mid)$ enters from phantom to non-phantom phase and stays in
it while the universe becomes matter dominated as EoS parameter
becomes zero for $z\rightarrow\infty$. The parameter
$\rho_{DE}^{(*)}$ represents the positive increment in its value as
a function of $z$.
\item The behavior of model V remains the same as that of the model IV.
However, the parameter $\omega_{DE}$ indicates a matter dominated
universe for both cases.
\end{itemize}
The comparison of our results with Bamba et al. (2011) is as
follows: For model I, $\omega_{DE}(z)$ stays in the phantom phase
for $p>0$ and non-phantom phase for $p<0$ in FRW universe, whereas
in BI universe, it shows opposite behavior. $\omega_{DE}(\mid
T/T_{0} \mid)$ does not cross phantom divide line in BI universe
while it crosses phantom divide line in FRW universe for models I
and II. $\omega_{DE}(z)$ gives the same behavior for both BI and FRW
in model II. The crossing of phantom divide line occurs in combined
model for both BI and FRW. In models IV and V, the universe turns
out to be matter dominated era in BI universe while these models
behave like the cosmological constant in FRW universe. We can
conclude from the above discussion that the DE component is
responsible for the accelerating expansion of the universe. This is
consistent with recent observations like SNIa and WMAP data
(Perlmutter et al. 1999; Knop et al. 2003; Riess et al. 1998).

\end{document}